# Near-Field Engineering of Silica-Shelled Plasmonic Nanoparticles toward Enhanced Near-Infrared Fluorescence

Eleonora G. Radaelli[a], Stavros Athanasiou[b], Robin E. Schäublin[c], Olivier J. F. Martin[b], Markus Niederberger[a]*, Fatemeh Kiani[a]*

[a] *Laboratory for Multifunctional Materials, Department of Materials, ETH Zurich, Vladimir-Prelog-Weg 5, Zürich, Switzerland*

[b] *Nanophotonics and Metrology Laboratory (NAM), Swiss Federal Institute of Technology Lausanne (EPFL), Lausanne 1015, Switzerland*

[c] *Scientific Center for Optical and Electron Microscopy, ETH Zurich, Otto-Stern-Weg 3, Zurich, Switzerland*

**Corresponding authors: Markus Niederberger (markus.niederberger@mat.ethz.ch) and Fatemeh Kiani (fatemeh.kiani@mat.ethz.ch)*

## Abstract

Near-infrared (NIR) fluorescent molecules often suffer from low quantum yields, limiting their applications in sensing and imaging. Their fluorescence can be enhanced by coupling with plasmonic nanoparticles exhibiting resonances in the same spectral region. However, this approach requires precise nanoscale control over the plasmonic near-field. Here, we combine scanning transmission electron microscopy with electron energy-loss spectroscopy (STEM-EELS) and boundary element electromagnetic simulations to engineer silica-coated anisotropic plasmonic nanoparticles. We synthesised three types of nanoparticles, gold nanorods (AuNRs), gold bipyramids (AuBPs), and silver nanorods (AgNRs), with tunable NIR plasmonic resonances. Silica serves as a dielectric spacer, enabling precise control over the distance between the nanoparticle surface and a potential molecular emitter. To assess their suitability for fluorophore attachment, we investigated the dependence of the electric near-field on silica shell thickness. High-resolution STEM-EELS mapping revealed that thin silica shells (4.4 nm for AuNRs and 3.1 nm for AuBPs) allow most of the near-field to extend beyond the dielectric layer. For AgNRs, a thicker shell of 7.0 nm was required to prevent oxidation while maintaining near-field accessibility. Electromagnetic simulations further evaluated the optimal shell thickness and the dependence of fluorescence enhancement on particle geometry, material, fluorophore internal quantum yield, and fluorophore–metal separation. Together, these results provide practical design guidelines for optimizing particle morphology and sub-10 nm silica shells in plasmonic nanostructures for fluorescence enhancement.

## Introduction

A localised surface plasmon is defined as the collective oscillation of the free electrons within a nanoparticle, which results in the generation of strongly confined electromagnetic fields in the vicinity of the nanoparticle surface. This collective motion occurs when the frequency of the incoming light matches a specific frequency, known as localised surface plasmon resonance (LSPR).[1-3] This resonance frequency can be influenced by the material and the geometry of the nanoparticle as well as the characteristics of its surrounding medium.[1-3] As a result, LSPRs can span from the ultraviolet up to the infrared.[1-4] Plasmonic nanoparticles (pNPs) with resonances in the near infrared (NIR) spectral region are of particular relevance. This is because many organic fluorescent molecules that emit in this region are attractive candidates for use in bioimaging and optical devices, but they usually suffer from poor photon emission.[5-7] In fact, the emission rate of fluorescent molecules is influenced by the electromagnetic local density of states (EMLDOS), which represents the probability of identifying an electromagnetic mode at a given point in space and energy.[8] Since plasmonic nanoparticles strongly and locally enhance the electromagnetic field, the EMLDOS increases. Therefore, coupling fluorophores to a plasmonic field increases their radiative decay rate, resulting in enhanced fluorescence emission. Since fluorescence is the product of an absorption and an emission contribution, enhancing at least one of them leads to plasmon-enhanced fluorescence (PEF).[8-12] To achieve fluorescence enhancement, three aspects are relevant: (i) the plasmonic field of the nanoparticle, (ii) the orientation and quantum yield (QY) of the fluorophore, and (iii) the spectral overlap between the emission of the dye with the plasmon band. In turn, these aspects are dependent on the material and geometry of the nanoparticle, as well as on intrinsic properties of the selected molecule.[12] In this context, anisotropic pNPs are especially useful because their morphology enables both strong field confinement and tunable plasmon resonances. For example, gold nanorods (AuNRs) and gold bipyramids (AuBPs) support longitudinal plasmon modes that can be shifted into the near infrared (NIR) by controlling their aspect ratio and can have highly localised electromagnetic hotspots depending on the sharpness of their tips.[13, 14] Silver nanorods (AgNRs) offer potentially stronger plasmonic enhancement compared to their gold counterparts but require sufficient surface protection due to their lower chemical stability.[1, 15, 16] Dielectric shells can be used to prevent the nanoparticles from being damaged and to control the spectral overlap between the dye emission and the plasmonic band. Particularly, silica ($SiO_2$) coatings are chemically stable and optically transparent. Their thickness can be tuned at the nanometre scale while still providing a versatile surface for subsequent fluorophore functionalization.[17-21] However, this approach introduces an inherent design trade-off: the coating must be sufficiently thick and homogeneous to physically separate the emitter from the pNP surface and stabilise the nanoparticle, but sufficiently thin to allow the plasmonic field to radiate through. In fact, plasmonic enhancement becomes negligible when the molecule is too far from the surface, as the electromagnetic field of pNPs decays with increasing distance from the particle surface. However, the direct attachment of molecules with moderate to high quantum yields to the surface of pNPs has been shown to quench fluorescence.[12, 22] Therefore, precise control over the separation between particle and molecule is necessary to achieve maximum fluorescence enhancement. The definition of an optimum shell thickness becomes particularly challenging for anisotropic pNPs, as their electromagnetic field is concentrated at specific locations such as tips, edges, and side surfaces.[23, 24] UV–vis–NIR spectroscopy is commonly used to identify resonance energies and changes in the dielectric environment following shell growth,[17, 21] but it cannot directly determine where the electromagnetic modes are localised or how far their fields extend through a nanometre-scale dielectric shell.

Scanning transmission electron microscopy combined with electron energy-loss spectroscopy (STEM-EELS) allows to overcome this limitation. EELS is a powerful method for studying the local excitation of surface plasmons with nanometer spatial resolution. Indeed, Garcia de Abajo and Kociak have demonstrated that the EELS signal is connected with the EMLDOS.[25, 26] Therefore, low-loss STEM-EELS enables the localization of longitudinal, transverse, and higher-order plasmon modes around individual nanoparticles.[26, 27] EELS has been widely employed to investigate plasmon modes in metallic nanostructures and core–shell systems,[23, 24, 28, 29] yet its potential as a design tool for controlling the $SiO_2$ spacer thickness in anisotropic plasmonic nanoparticles is largely unexplored. The combination of experimental STEM-EELS maps with surface integral equations (SIE) is a particularly insightful approach, as it connects the experimentally observed mode distributions with the theoretical interpretation of the near fields, ultimately giving information on the electromagnetic environment experienced by a fluorophore.

In this work, we establish an EELS-guided approach for engineering silica-coated anisotropic plasmonic nanoparticles for fluorescence enhancement. Anisotropic plasmonic nanoparticles, namely AuNRs, AuBPs, and AgNRs, were systematically coated with a thin silica dielectric layer with varying thicknesses (from 2 to 20 nm). The spatial distribution of their plasmonic modes is investigated by STEM-EELS, while boundary-element electromagnetic calculations provide a direct comparison with the experimental field distributions. The simulations are then extended to evaluate potential fluorescence enhancement as a function of nanoparticle morphology, metal composition, fluorophore quantum yield, and fluorophore–metal distance. This combined experimental–computational approach identifies distinct design regimes for the three nanoparticle systems and, in the case of AgNRs, reveals a trade-off between maximising near-field accessibility and providing sufficient protection against degradation. The proposed framework establishes a method for selecting the most suitable material and morphology for metal-silica core-shell nanoparticles in the context of fluorescence enhancement. Moreover, these results set the basis for potential fluorescence enhancement studies of NIR fluorophores-metallic nanoparticles systems.

## Results and Discussion

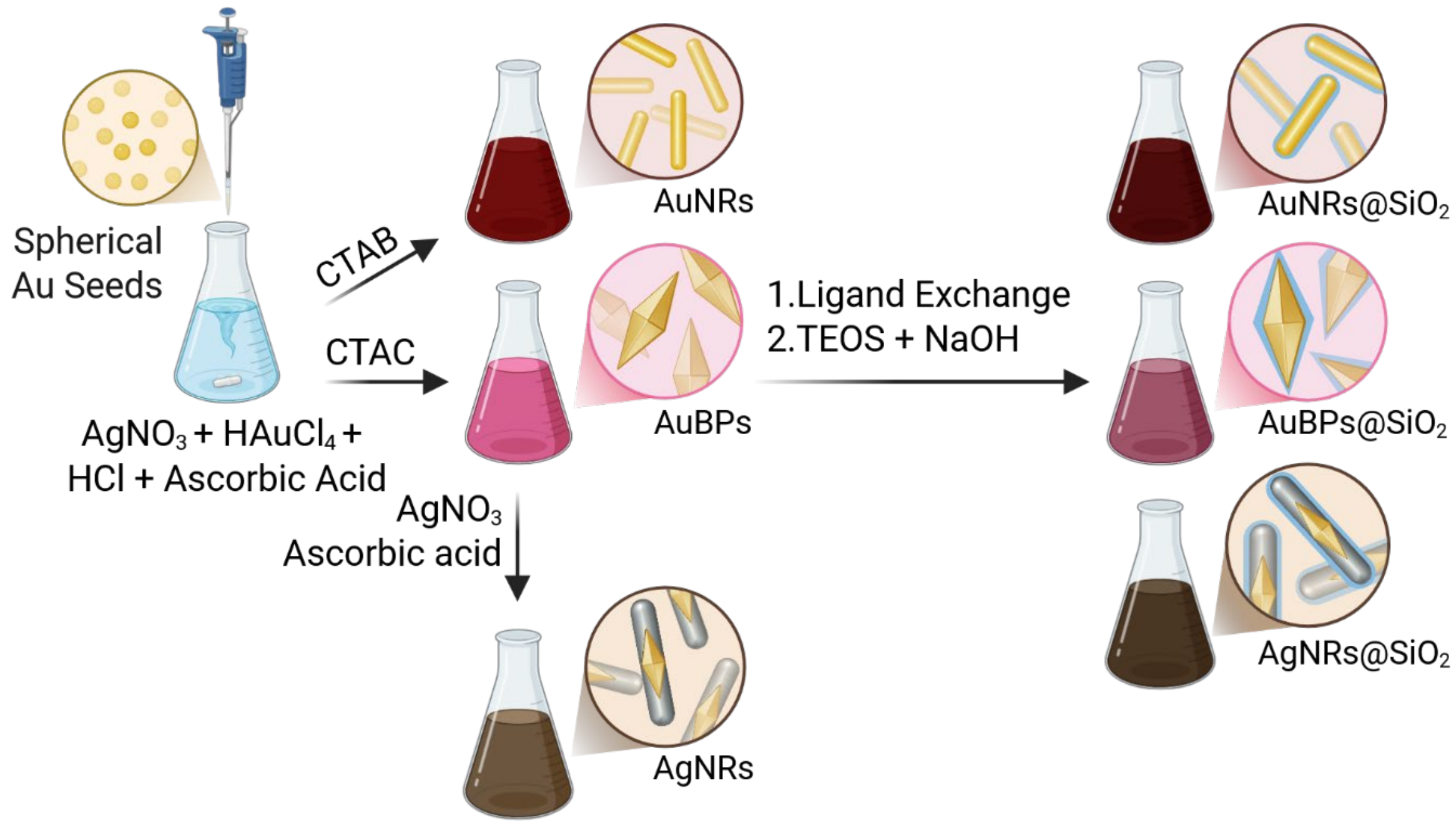

**Figure 1:** Reaction scheme of the multi-step synthesis of silica shelled AuNRs, AuBPs and AgNRs (here generally referred to as pNPs@$SiO_2$ with pNPs substituted by the specific nanoparticle system).

In this study, we propose an EELS-guided approach for engineering silica-coated anisotropic plasmonic nanoparticles (pNPs) for plasmon enhanced fluorescence (PEF). To be able to study the plasmonic field of such silica ($SiO_2$) shelled pNPs, we first synthesised the pNPs: gold nanorods (AuNRs), gold bipyramids (AuBPs) and silver nanorods (AgNRs) (**Fig.**1). The nanoparticle synthesis followed two distinct multistep protocols previously reported by González-Rubio and Sánchez-Iglesias, according to the following rationale: (i) the preparation of either single crystalline or penta-twinned gold spherical seeds, (ii) the production of anisotropic seeds, and (iii) the growth of the anisotropic seeds into larger particles.[13, 14] González-Rubio *et al.* showed that the quality of the nanocrystalline seeds strongly influences the subsequent production of anisotropic nanoparticles: the spherical seeds need to be around 1-2 nm and freshly prepared as they are only stable in a short timeframe (a few hours).[13, 30] Therefore, controlling this step is crucial for the overall process. Gold spherical seeds are prepared by rapidly injecting a strong reducing agent (sodium borohydride, $NaBH_4$) into a solution of partially reduced gold (III) chloride trihydrate ($HAuCl_4 \cdot 3\ H_2O$). Specifically, before the $NaBH_4$ addition, the gold (III) complex is reduced to Au (I) with freshly prepared ascorbic acid (light-sensitive), to improve the nanocrystal size control. Moreover, surfactants also play an active role in this first step: while hexadecyltrimethylammonium bromide (CTAB) is sufficient to produce AuNRs with the desired morphology, hexadecyltrimethylammonium chloride (CTAC) is necessary to induce the formation of twinned defects which are fundamental to obtain AuBPs. The spherical seeds are then further grown into anisotropic gold rods or penta-twinned bipyramids. This process is enabled by silver nitrate ($AgNO_3$). In fact, $Ag^+$ ions are essential for reducing gold ions locally at the AuNRs tips and stabilising lateral facets in both AuNRs and AuBPs. Importantly, $AgNO_3$ in solution is not light stable and is therefore prepared right before use. In this step, hydrochloric acid (HCl) is also added to the solution to enable the tuning of the plasmonic band. The growth of AuNRs and AuBPs is then initiated upon addition of the spherical gold seeds to this precursor-containing solution. Furthermore, the obtained AuBPs are used as anisotropic seeds to grow AgNRs by inducing the local reduction of $AgNO_3$ on the AuBPs with ascorbic acid.[15] After synthesizing plasmonic nanoparticles, we coat them with $SiO_2$ using a modified Stöber process.[17] This involves the priming of the surface with (3-mercaptopropyl) trimethoxysilane and adjusting the pH to 9 to ensure homogeneous shell growth. Before each coating, it was fundamental to determine the starting pH as well as the number of pNPs and their surface area to ensure that the quantity of $SiO_2$ precursor was comparable across batches. The produced silica-coated plasmonic nanoparticles (pNP@$SiO_2$) are then screened by UV-vis-NIR spectroscopy and scanning transmission electron microscopy (STEM). Specifically, UV-vis-NIR spectroscopy and STEM analysis are carried out to ensure shape purity, namely the absence of particles with different morphologies from the desired ones, and the presence of a homogenous coating on the pNPs throughout each batch.

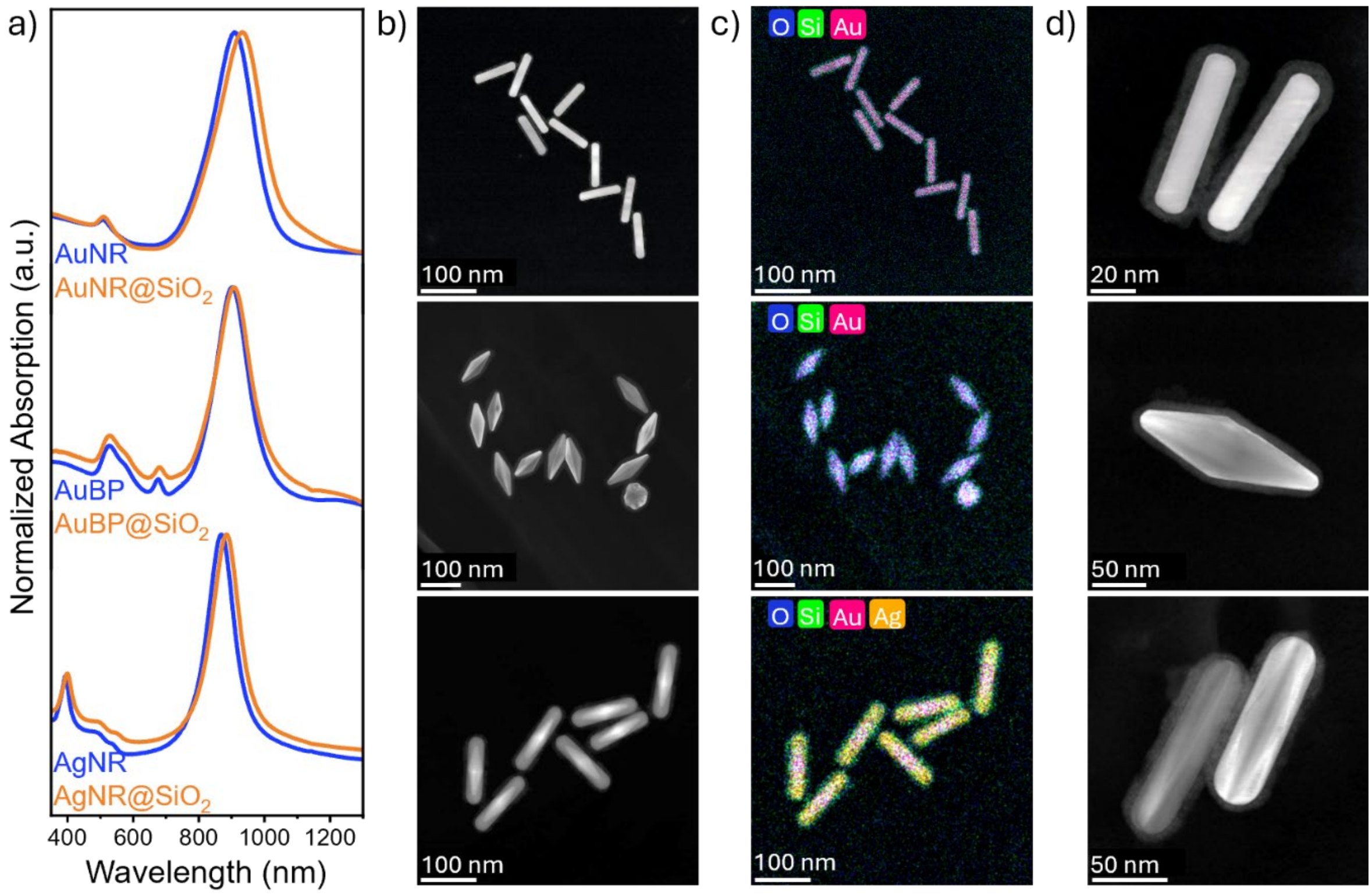

**Figure 2:** (a) UV-vis-NIR normalized absorption spectra in water of AuNRs, AuBPs and AgNRs, before and after coating with silica. The absorption is normalized to the main dipolar L-mode. (b) Low-magnification dark field (DF)-STEM images and (c) EDX O-K, SI-K, Au-L and Ag-L elemental maps of shelled AuNRs, AuBPs and AgNRs. (d) High-magnification DF-STEM images of shelled pNPs where the coating is clearly distinguishable.

The extinction of each pNP before and after coating with silica was studied by UV-vis-NIR spectroscopy (**Fig.** 2a). For all anisotropic nanoparticles, at least two modes are observed: a transverse mode (T-mode) in the visible range and a dipolar longitudinal mode (L-mode) at longer wavelengths. Additionally, higher-order multipoles arise when the material or its geometry changes. For example, increasing the aspect ratio of an anisotropic nanoparticle causes the appearance of a secondary longitudinal mode, together with the shift of the dipolar L-mode.[1] The extinction of AuNRs is characterised by a T-mode (ca. 510 nm) and a dipolar L-mode (in this case at ca. 907 nm).[1] Moreover, the complete absence of secondary maxima indicates that high shape purity was achieved. In the spectra of AuBPs, besides the T-mode (ca. 525 nm) and the first dipolar L-mode (ca. 900 nm), a second dipolar mode arises at ca. 675 nm, due to the high aspect ratio between the length and the diameter of the nanoparticle.[14, 15] The line shape of AuBPs also reveals a secondary maxima at ca. 570 nm. This feature is absent in previous reports and it reveals that a small portion of side-products could still be present.[14] The extinction of AgNRs is characterised by four bands at ca. 868 nm, 536 nm, 495 nm and 390 nm, previously attributed to the dipolar L-mode, the quadrupolar L-mode, the octupolar L-mode, and dipolar T-mode, respectively.[15] Furthermore, most particles have the desired shape as further verified by low-magnification STEM analysis (**Fig.** 2b). In detail, the average lengths and width for AuNRs and AgNRs are 77 ± 7 nm and 16 ± 2 nm, and 123.3 ± 8 nm and 33.7 ± 6 nm, respectively. The standard deviation of the average lengths is within 10% and each nanorod is uniform in diameter along the length direction. For AuBPs, the average length and waist width are 190.4 ± 7 nm and 63.5 ± 4 nm, making AuBPs and AgNRs more than twice the size of AuNRs. After the pNPs are coated with silica, the modes characterizing the line shapes appear at slightly red-shifted wavelengths (**Fig.** 2a). This bathochromic shift of ca. 20 nm for AuNRs, 7 nm for AuBPs, and 15 nm for AgNRs is to be expected, as it is the effect of the different dielectric environment on the plasmonic field of the pNPs. Indeed, when the pNPs are uncoated, the dielectric constant is that of the solvent in which they are analysed (in this case, water). After coating, however, the plasmonic field is red-shifted due to the higher dielectric constant of the $SiO_2$ shell compared to that of water. The presence of a homogeneous silica shell was further verified by energy-dispersive X-ray (EDX) spectroscopy elemental mapping (**Fig.** 2c) which unequivocably identifies the thin shell to be made of silicon and oxygen. Moreover, high-magnification STEM analysis (**Fig.** 2d) was used to evaluate the control over the shell quality. The average shell thicknesses measured were calculated to be 5.4 ± 1.0 nm, 6.75 ± 0.9 nm and 6.4 ± 1.5 nm for AuNRs, AuBPs, and AgNRs, respectively. The relatively high standard deviation in shell thickness is mostly caused by variations in the coating thickness around each nanoparticle. Indeed, for AuNRs and AgNRs, the shells are slightly thinner at the tips than on the sides, whereas for AuBPs, the coatings tend to be thicker at the tips. We hypothesise that this is due to the surface primer and silica precursor having a different affinity for certain facets, or to the surfactant (CTAB) being inhomogeneously distributed on the surface. We also observed that very thin shells (below 3 nm) tend to be less uniform. Therefore, we verified by UV-vis-NIR spectroscopy and STEM-EDX analysis that we obtain pNPs with high shape purity and sufficiently homogenous $SiO_2$ coatings.

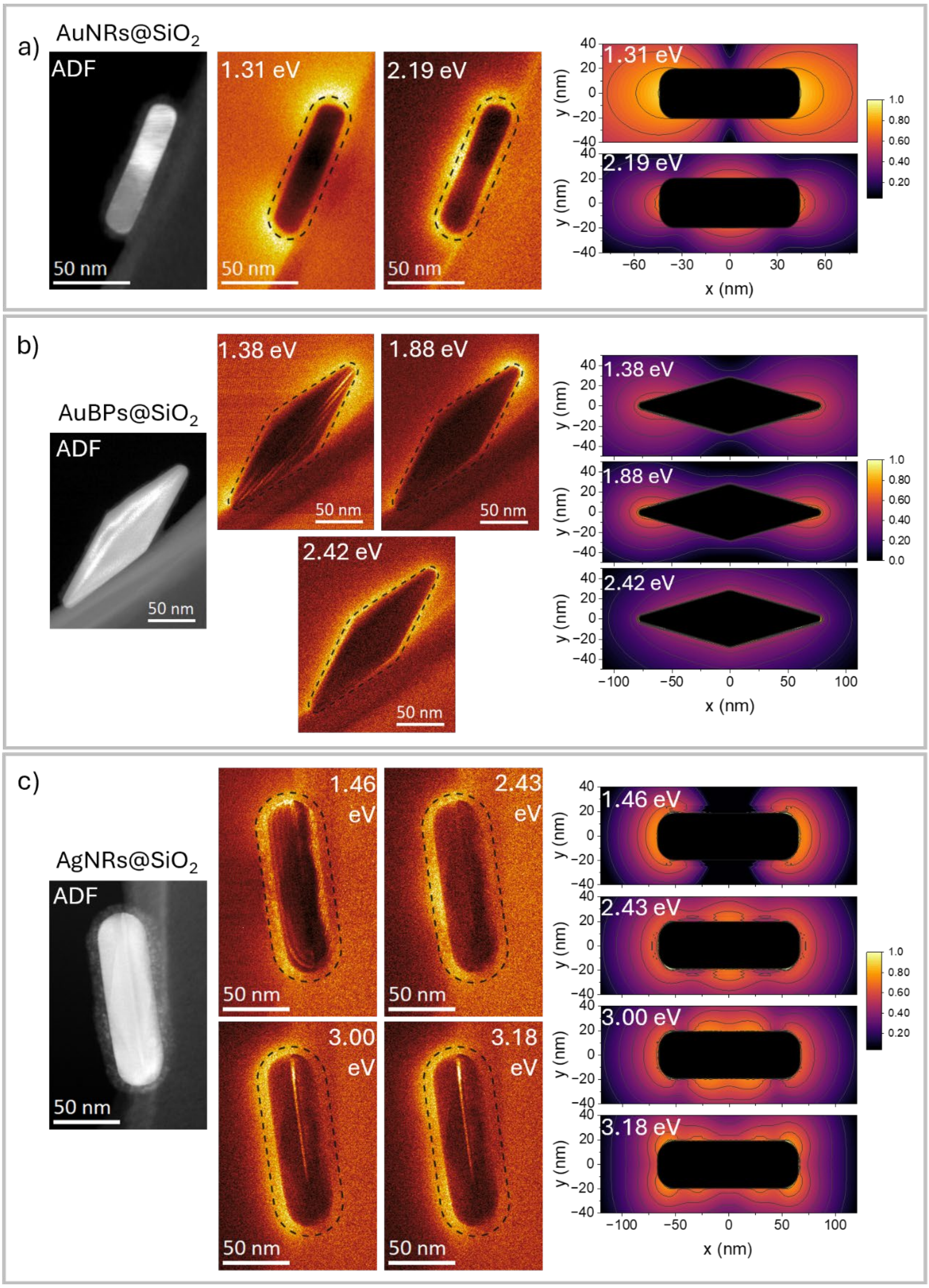

**Figure 3:** ADF-STEM images, STEM-EELS spectrum images and respective boundary element electromagnetic calculations of a) AuNRs@$SiO_2$, b) AuBPs@$SiO_2$, and c) AgNRs@$SiO_2$.

To investigate the effect of the shelling on the near-field of pNPs@$SiO_2$, STEM-EELS spectrum images of the coated pNPs on a lacey carbon support were acquired with the respective annular dark-field (ADF) STEM images (**Fig.** 3, left). At the same time, boundary elements electromagnetic calculations (**Fig.** 3, right) were carried out to propose a physical interpretation of the STEM-EELS data and exclude the presence of artefacts in the maps. Specifically, the STEM-EELS spectrum images are acquired in vacuum (except for the fraction of the pNP in contact with the carbon support) and show the location of the dipolar and higher-order multipolar modes of $SiO_2$ shelled AuNRs, AuBPs, and AgNRs.

Given the dielectric nature of $SiO_2$, we expect that the shell thickness affects the distance within which the near field of the pNPs decays. Prior to any further considerations, we would like to specify that, in each STEM-EELS spectrum image that we studied, the focus is on the fraction of pNP that is fully in vacuum. This is due to the carbon support used in the experiment potentially affecting the plasmonic response of the nanoparticles. Figure 3a (left) shows the STEM image acquired via ADF detector and the respective near field spectrum images of a AuNR with a 4.4 nm silica shell. The dipolar L-mode is identified at 1.305 ± 0.240 eV when averaging the EELS spectra of the area around the tip of the AuNR, and the T-mode is identified at 2.190 ± 0.240 eV when averaging the spectra at the side of the AuNR. This observation agrees with previous reports of the near field of bare AuNRs[23] and with the far-field results from UV-vis-NIR spectroscopy. Despite the silica coating, the dipolar L-mode field is clearly localised and more intense outside of the shell rather than inside it. In **Fig.** 3b (left), we show the ADF image and the near field of a AuBP with a 3.1 nm silica shell. In this case, the main dipolar L-mode is found around the tip of the pNP at 1.380 ± 0.240 eV, while the second dipolar L-mode and the T-mode are found at 1.882 ± 0.240 eV and at 2.430 ± 0.240 eV, when averaging the EELS spectra around the side of the AuBP. This is also comparable with the low-loss STEM-EELS findings on uncoated AuBPs,[31] as well as with the far-field observations in **Fig.** 2. The field corresponding to dipolar L-mode is very intense and localised outside the silica coating, in a small area around the tips of the shelled AuBP. Finally, in **Fig.** 3c (left), we report the near field of a AgNR with a 7.0 nm silica shell. The dipolar L-mode and the quadrupolar L-mode are strongest at the tip of the pNP, at 1.455 ± 0.240 eV and at 2.430 ± 0.240 eV, respectively. The dipolar T-mode is strongest when averaging the EELS spectra around the side of the pNP at 3.180 ± 0.240 eV, while an additional mode was identified when averaging the spectra in the region between the side and the tip of the AgNR, at approximately 3.000 ± 0.240 eV. To the best of our knowledge, comparable geometries have not previously been investigated using low-loss STEM-EELS, preventing a direct comparison with the literature. Despite slight shifts in energy, the near-field plasmon bands identified by STEM-EELS mapping with the far-field features described in the UV-vis-NIR spectroscopy section are consistent. The dipolar L-mode (ca 868 nm, 1.42 eV), of the T-mode (ca 390 nm, 3.17 eV), as well as of the quadrupolar L-mode (ca 536 nm, 2.31 eV) of AgNRs identified by far field extinction (**Fig.** 2a) are shifted to higher energies in the STEM-EELS analysis. The additional mode at approximately 3.000 ± 0.240 eV, does not seem to directly correspond to the octupolar L-mode which should be expected according to UV-vis-NIR spectroscopy (**Fig.** 2a) at roughly 495 nm (ca. 2.51 eV) and is very similar in shape to the mode at 3.18 eV. On the one hand, slight shifts at lower energies (close to 1 eV) could be explained by the absence of a monochromator in our setup. On the other hand, we hypothesize that a non-perpendicular orientation of the electron beam to the pNP, as previously shown by Aguilar *et al.* for AuNRs, could explain the bigger discrepancies in energy.[23] Moreover, the 7 nm coating, when compared to the thinner shells of AuNRs and AuBPs, seems to partially reduce the strength of the plasmonic near-field. Indeed, the local intensity of the dipolar L-mode of AgNRs is weaker outside the shell compared to the one inside the shell. Nonetheless, this thickness was necessary as we observed that thinner shells sometimes resulted in non-homogeneous coatings, which in turn caused the nanoparticles to chemically degrade and leak silver over time. This trade-off in shell thickness to optimise both chemical stability and near-field strength was not an issue for AuNRs and AuBPs, as gold is intrinsically more stable than silver.[1] Importantly, the localisation of the near field is less pronounced in the shelled AuNRs and AgNRs than in AuBPs. This implies that, if the surface of a nanorod was fully coated with fluorophores, a greater portion of molecules would be affected by its near field.

To further verify the attribution of the multipolar modes of AuNRs@$SiO_2$, AuBPs@$SiO_2$ and AgNRs@$SiO_2$, simulations based on surface integral equations (SIE) were carried out. Specifically, we performed frequency-domain electromagnetic simulations using the boundary-element method (BEM) of each particle in vacuum. For simplicity and computational efficiency, we restricted our analysis to nonpenetrating electron trajectories. Under these conditions, we treated the electron in the nonrecoil approximation as moving at constant velocity along a straight trajectory $\mathbf{r}_e(t) = (\mathbf{b}, vt)$, where **b** is the impact parameter in the plane perpendicular to the trajectory. The EELS probability was therefore evaluated from the induced electromagnetic Green tensor by representing the electron current as a coherent distribution of longitudinal electric-dipole sources along its trajectory, repeating this calculation over each experimental scan positions.[25, 32] The resulting simulated EELS maps in **Fig.** 3 (right) display the calculated loss probability $\Gamma(\mathbf{b}, \omega)$ at selected energy losses $\hbar\omega$ (indicated at the top of each figure in eV). First, it's important to note that in each simulation we have omitted the carbon support used in the experiment to isolate the intrinsic plasmonic response of the nanoparticles. By this approximation, we exclude substrate-induced changes in the resonance energies, linewidths, and modal intensities. When comparing the SIE simulations with the STEM-EELS spectrum images (**Fig.** 3) of shelled AuNRs and AuBPs, we observed that there is a good agreement between theory and experiment. This observation confirms the respective attribution of longitudinal and transverse modes for both systems. As previously observed by comparing the results of UV-vis-NIR spectroscopy (**Fig.** 2) and near-field STEM-EELS images (**Fig.** 3, left), the attribution of the longitudinal and transverse dipolar modes, as well as the longitudinal quadrupolar mode, is consistent for AgNRs@$SiO_2$. In contrast, the simulated map at 3.00 eV and the corresponding STEM-EELS image do not coincide. We hypothesize that this is due to absence of a monochromator and the lack of control over further parameters, such as the orientation of the nanoparticle to the electron beam,[23] during the STEM-EELS analysis.

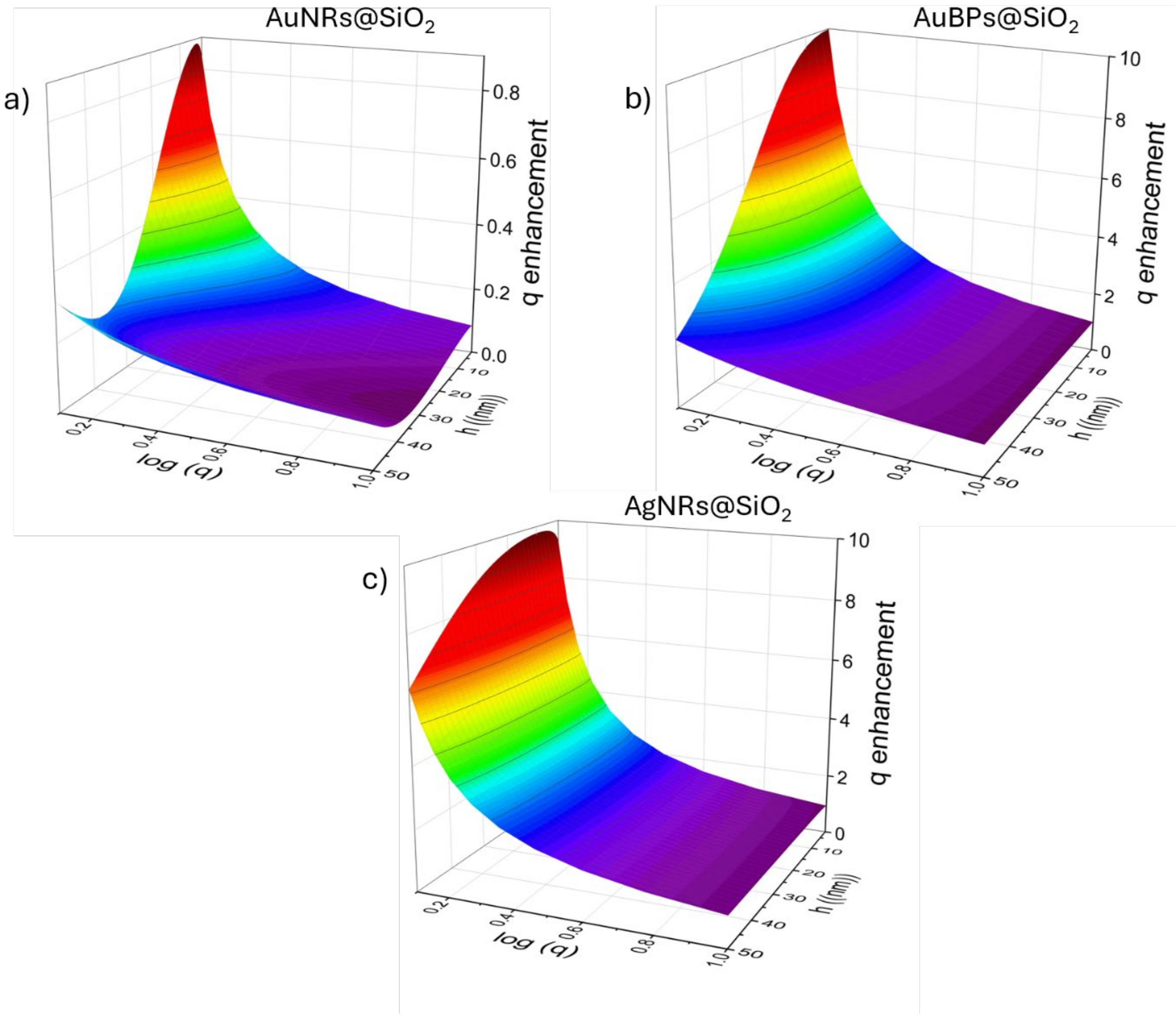

**Figure 4:** 3-dimensional boundary element electromagnetic calculations of the quantum-yield enhancement for the three geometries: a) AuNRs@$SiO_2$, b) AuBPs@$SiO_2$, and c) AgNRs@$SiO_2$. Specifically, the quantum-yield enhancement (vertical axis) is defined as the nanoparticle-modified quantum yield normalized to the intrinsic value and expressed as a function of the emitter–surface separation (h) and the intrinsic quantum yield (q) in logarithm scale (indicated by log(q)). We assume a homogeneous background medium of dimethylformamide (DMF), with a refractive index of (n = 1.4305).

With this work we also want to establish a framework for identifying the conditions under which AuNRs, AuBPs and AgNRs are best suited for plasmon-enhanced fluorescence applications. EELS measurements provide spatially resolved information about the plasmonic excitations of these nanoparticles but they do not directly yield the local dipole-projected photonic density of states that governs the spontaneous emission rate of a localised emitter.[33] We therefore complement the EELS measurements and calculations with supplementary calculations of the modified spontaneous emission rate and its radiative and nonradiative contributions (see Methods). Our analysis focuses exclusively on modifications to the emission process. Specifically, we assume that the excitation rate is unaffected by the nanoparticle, as may be approximately realized through optical excitation sufficiently far from the plasmonic resonances or through an appropriate non-optical excitation mechanism. The quantity of interest is the fluorescence quantum yield of a generic organic molecule, defined as the fraction of emitter excitations that decay through radiative channels. For each geometry, a generic emitter is positioned near the nanoparticle tip, and the quantum-yield enhancement is calculated and plotted (**Fig.** 4). Under the conditions considered, AuBPs and AgNRs (**Fig.** 4b-c) provide quantum-yield enhancements approximately ten times greater than those obtained with AuNRs (**Fig.** 4a). Specifically, the improved performance of AgNRs, which contain a gold bipyramid seed, is consistent with the lower optical losses of silver at the relevant wavelengths. For AuBPs, the sharper tips promote stronger field confinement and emitter–plasmon coupling than in the AuNR geometry, which strongly favours enhanced radiative emission. The more confined response of AuBPs is also consistent with the faster decrease in quantum-yield enhancement with increasing separation beyond the optimum. For AuNRs and AuBPs, the optima occur at short separations, approximately at 4 nm (**Fig.** 4a) and below 2 nm (**Fig.** 4b), while AgNRs support substantial enhancement over a broader range of separations, with an optimum enhancement corresponding to shells that are 7-12 nm thick. In general, we observed that emitters with very low intrinsic quantum yields can benefit more substantially from radiative-rate enhancement, provided that this term sufficiently outweighs the additional nanoparticle-induced nonradiative decay. For emitters with intrinsic quantum yields close to unity, the scope for further enhancement is limited, while additional nonradiative decay through absorption in the nanoparticle can reduce the quantum yield. For an intrinsic quantum yield of exactly unity, no quantum-yield enhancement is possible.

Based on these observations, we establish in which conditions shelled AuNRs, AuBPs and AgNRs are best employed for maximising the fluorescence of organic fluorophores, as a function of silica thickness and the distribution of the emitters on the nanoparticle surface. AuNRs have the advantage of being much smaller than AuBPs and AgNRs with plasmon bands at comparable energies while offering modest enhancements. This makes them ideal for proof-of-concept studies where smaller size outweighs the need for maximum local enhancement. For these cases, a thin shell of approximately 4 nm is best, and the dye can be uniformly distributed throughout the surface of the pNPs. AuBPs offer the maximum local enhancement among all the tested particles, making them ideal when performance is a critical requirement. Given the strong localization of the field, precise chemical attachment to the tip region is recommended to best harness this advantage. Since the enhancement stems from extreme field confinement at the AuBPs tips and drops off rapidly beyond 5 nm but shells become less uniform when they are thinner than 2 nm, a very thin shell of 3 nm is suggested. Finally, AgNRs also show strong quantum yield enhancement with the advantage of being less susceptible to nanometer variations in shell thickness (a range of 7 to 12 nm coating thickness offers comparable enhancements) than AuNRs and AuBPs. Furthermore, the near-field of AgNRs is more spread around the entire nanoparticle surface, rather than in a very small region as for AuBPs. This makes coated AgNRs optimal for applications where strong enhancement is required and bypasses the need for a locally precise attachment. However, AuNRs and AuBPs are more suitable for applications where long-term performance is relevant, due to the better chemical stability of gold over silver. Importantly, independently of the materials, both tips of nanorods and bipyramids favour enhanced radiative emission (although to different extents). However, the resulting quantum yield enhancement of any emitter depends on the balance between radiative decay and nonradiative dissipation, rather than on field confinement alone. Therefore, the overall increase in quantum yield strongly depends on the characteristics of the organic fluorophores. For this reason, this study poses the basis for potential experimental validation of the coupling of these pNPs@$SiO_2$ with NIR-active fluorescent organic molecules.

## Conclusion

In this study we focus on the nanometre-scale engineering of silica-coated gold nanorods, gold bipyramids and silver nanorods (AuNRs, AuBPs, and AgNRs) with the aim of defining the localization of the plasmonic field as well as the optimal silica ($SiO_2$) shell thickness for plasmonic field enhancement. By using a combination of STEM-EELS and boundary element electromagnetic calculations, we establish a framework for selecting $SiO_2$-shelled plasmonic nanoparticles (pNPs) based on their potential fluorescence enhancement. Our combined experimental-computational approach highlights that AuBPs allow for an ten-fold strong enhancement with a shell thickness of 3 nm. To leverage their strongly confined near-field, molecules should be strategically positioned at the AuBPs tips. AgNRs provide a nine-fold high enhancement compared to AuNRs. However, their performance relies on the quality of their silica shell. Consequently, thicker coatings of 7 nm or more are required to ensure long-term chemical stability. While offering more modest enhancements, AuNRs are much smaller in size and active at resonance wavelengths comparable to those of AuBPs and AgNRs. With an optimal 4 nm shell and uniform dye distribution, they are ideal for proof-of-concept studies. Furthermore, we observed that emitters with very low intrinsic quantum yields benefit the most from radiative-rate enhancement. By pairing appropriate fluorophores with optimized shell thickness and pNPs material and geometry, the nonradiative dissipation is minimized to optimize performance in each scenario. Ultimately, this STEM-EELS and electromagnetic calculations combined workflow is readily transferable to other anisotropic plasmonic core–shell systems where near field control at the nanometre scale is critical. The herein proposed framework provides the foundation for establishing practical guidelines to facilitate the selection of pNPs for fluorescence enhancement and lays the foundation for potential experimental coupling of AuNRs, AuBPs, and AgNRs with NIR-fluorescent organic molecules.

## Methods

### *Chemicals*

All chemicals were used as received without further purification. Hexadecyltrimethylammonium bromide (BioXtra, >99%, CTAB), Hexadecyltrimethylammonium chloride (>=98.0%, CTAC), 1-decanol (99%), gold (III) chloride trihydrate (> 99.9% trace metal basis), sodium borohydride (99%), silver nitrate (puriss. p.a., >99.5%; for the synthesis of AuNRs), L-ascorbic acid (>99%), ethanol (absolute >99.8% for analysis), citric acid (99 %), 3-mercapto propyltrimethoxysilane (95%, MPTMS), tetraethoxysilane (>99.0% GC, TEOS), sodium hydroxide (stock solution 0.1N Ampulle, diluted in a 1L volumetric flask), and dimethylformamide (>99% extra pure), were purchased from Sigma-Aldrich. Hydrochloric acid (37%, AnalaR, NORMAPUR), hydrochloric acid (1M solution), methanol (>99.8%, analytical reagent), acetonitrile (AnalaR NORMAPUR), dimethyl sulfoxide (99.9% for spectroscopy) were purchased from VWR Chemicals. Silver nitrate (>99.995%, metal basis) used for AuBPs and AgNRs was purchased from abcr GmbH. Finally, tetrahydrofurane (LiChrosolv) was purchased from Merck. Ultrapure water (18.2 MΩ cm at 25 °C) was obtained using a Merck Millipore Synergy purifier.

### *Synthesis of gold nanorods*

The AuNRs were synthesised using a three-step seed-mediated method previously described by Gonzàlez-Rubio[13] including: (i) the preparation of small spherical seeds, (ii) the growth of anisotropic seeds and (iii) the further growth of the elongated seeds into anisotropic particles. To synthesise 1-2 nm gold seeds, a surfactant solution was first prepared in a 500 ml Erlenmeyer flask by mixing 9.111 g CTAB (50 mM) and 1.068 g 1-decanol (13.5 mM) with 500 g of warm MilliQ water (stirring at ~60 °C for 1 h or 30 °C overnight and cooling it down to 25 °C before use). Then, 200 µL of a 0.05 M $HAuCl_4$ solution were added to 20 mL of the previously prepared surfactant solution in a 50 ml glass beaker under stirring at 400 rpm at 25 °C. To this orange solution 100 µL of a 0.1 M ascorbic acid solution are added, causing the solution to turn colourless. After 2-3 min, 800 µL of freshly prepared 0.02 M $NaBH_4$ solution were added under vigorous stirring (1400 rpm), causing the solution to turn instantaneously to a brown colour. The reaction was then left to age at 25 °C for 1 hour. Secondly, to obtain small anisotropic seeds, 230 mL of the previously prepared surfactant solution (50 mM CTAB, 13.5 mM 1-decanol) were mixed with 1.84 mL of a 0.01 M $AgNO_3$ solution and 2.3 mL of a 0.05 M $HAuCl_4$ solution in a 500 mL Erlenmeyer flask. 25.45 mL of this mixture was then transferred into a 50 mL vial and 324 µL of a 0.1 M ascorbic acid solution were added under stirring at 450 rpm. The resulting solution was divided into five 10 mL vials (5.155 ml of solution in each vial), and a certain volume (250 µL , 300 µL, 350 µL , 400 µL, 450 µL) of 1 M HCl followed by 300 µL of the gold seed solution were added to each vial under stirring at 400 rpm. The mixtures were left to mix and after 1 hour, after which the UV-vis spectrum of each sample was recorded to determine the optimal concentration of HCl. Once the optimum amount of HCl has been found, the synthesis was scaled up. Respectively, 2.664 mL of 0.1 M ascorbic acid, 12.299 mL of 1 M HCl (considering for example that 300 μL to be the optimum value of HCl), and 12.299 mL of spherical gold seeds were added in this order to 208.285 mL of leftover $HAuCl_4$, $AgNO_3$ and surfactant mixture in a 500 mL Erlenmeyer flask under stirring (450 rpm) at 25 °C. The solution was then left to stir overnight, leading to the growth of anisotropic gold rods, recognizable by a colour change to dark brown-grey. The solution was then centrifuged 3 times at 15000 rpm (20100 rcf, Universal 320, Hettlich Zentrifugen) for 45 min in 2 mL tubes. After each centrifugation, the surnatant was removed and the precipitate was redispersed with 14 mL of a 10 mM CTAB solution. For the synthesis of gold nanorods (aiming for a LSPR wavelength of 925 nm), 7.289 g of CTAB (50 mM) and 696.4 mg (11 mM) of n-decanol are solubilized in 400 mL of warm ultrapure water in a 500 mL Erlenmeyer under stirring (300-500 rpm) at 28 °C. Then, 8.0 mL of 0.01 M $AgNO_3$ solution, 4.0 mL of 0.05 $HAuCl_4$ solution, 3.2 mL of 0.1 M ascorbic acid solution and 18 mL of 1 M HCl were added the solution, followed by 2 mL of 4.65mM anisotropic gold seeds. This solution was left stirring for at least 5 hours or until the desired wavelength was reached. The gold nanorod solution was then centrifuged 3 times under 9000 rpm (8960 rcf, Universal 320, Hettlich Zentrifugen) for 45 min in 50 mL tubes, substituting the surnatant with 10 mM CTAB in MilliQ water each time. We therefore assumed that the final nanorod solutions have a CTAB concentration of 10mM.

***Synthesis of gold bypiramids and silver nanorods***

The synthesis of gold bipyramids and silver nanorods follows two previously reported reports by Sánchez-Iglesias *et al.* and Zhuo *et al.*.[14, 16] For gold bipyramids, the synthesis is articulated in two steps: (i) the production of penta-twinned gold seeds, and (ii) the growth of the seeds into bipyramids. To synthesize penta-twinned gold seeds, in a clean 50 ml glass vial 5 mL of aqueous CTAC (0.1 M) is prepared. To this solution, under stirring at 250 rpm with a 6x15 mm stir bar, 2.5 mL of $HAuCl_4$ (1 mM) and 2.5 mL of citric acid (0.02 M) were added. Then a fresh $NaBH_4$ solution (0.025 M) was newly prepared. 0.25 mL of this solution were added under vigorous stirring (1000 rpm) to the mixture, which instantly turned dark brown. The solution was left to stir for 2 minutes, after which the vial was placed in water bath at 80 °C and kept under mild stirring (250 rpm) for 2 hours. The seeds were then allowed to cool down to room temperature and used without further purification. After the penta-twinned seeds were produced, gold bipyramids were then synthesised by seed-mediated growth. Under stirring at 250 rpm, in a 30 °C water bath, 80 mL aqueous CTAB (0.1 M) were mixed in this order 4 mL $HAuCl_4$ (0.01 M), 0.8 mL $AgNO_3$ (0.01 M), 1.6 mL HCl (1 M), and 0.64 mL ascorbic acid (0.1 M). Under vigorous stirring (1000 rpm), 2.0 mL of the thermally aged Au seeds were injected in the solution and then maintained at 30 °C for at least 4 hours under mild stirring (250 rpm). During this time the solution turned dark purple. The obtained gold bipyramids were then purified by first centrifuging them at 1000 rpm for 15 min in 2 mL vials. The supernatant was then discarded, exchanged with 2 mL CTAC (0.01 M) and the solution was centrifuged again at 1000 rpm (89 rcf, Universal 320, Hettlich Zentrifugen) for 15 min. The centrifugation was repeated a total of three times, then the pellet was redispersed in 40 mL of CTAC (0.01 M). Purified AuBPs were used for the anisotropic templated growth of silver nanorods. To the 40 mL of gold bipyramids dispersed in CTAC (0.01 M) 8 mL of $AgNO_3$ (0.01 M) and 4 mL ascorbic acid (0.1 M) were added at room temperature without stirring. The mixture was then placed in a 60 °C water bath and kept under stirring (120 rpm) for 2 hours to ensure silver nanorod growth, to which corresponded a colour change to orange-brown.

***Synthesis of silica-coated gold nanorods***

The gold nanorods were coated using an adapted version of the synthesis reported by Pellas et al. [17] The purified gold nanorod solution was first diluted with Milli-Q water to achieve a concentration of 2.8 mM of Au (0) (calculated from their absorbance at 400 nm)[34], a surface area comparable to that of the reference paper (calculated from the data obtained by TEM analysis) and a CTAB concentration of ~ 0.5-0.9 mM. For this purpose, the concentrated AuNR solution was diluted with Milli-Q water in a 5 mL vial under stirring (~ 400 rpm). Whenever the dilution would lead to a CTAB concentration of less than 0.5 mM, a calculated amount of 10 mM CTAB was added to reach the threshold value. Moreover, the pH was measured with a pH meter and tuned to 4 by adding the corresponding amount of 0.1 M HCl. Then, 5.74 µL of a 100 mM MPTMS solution in ethanol was added to 700 µL of diluted gold solution and left under stirring (400 rpm) at 25 °C for three hours. The stirring speed was then increased to 800 rpm, after which 17.36 µL of a methanolic 20 vol.% TEOS solution was added to the vial, and the reaction was left to stir for an hour. Finally, 8 µL of 10 mM NaOH were injected after reducing the stirring speed to 400 rpm. The reaction mixture was stirred for a further 20 minutes, after which it was left to react undisturbed for at least 20 hours. The samples were then purified by high-speed centrifugation at 9000 rpm (8960 g, Universal 320, Hettlich Zentrifugen) for 15 minutes in 15 mL centrifuge tube diluting the dispersions with a total of 10mL of DMF. Then the samples were centrifuged at, and the supernatant was removed. Centrifuging was repeated twice more. Finally, the particles were redispersed with 2 mL of DMF.

***Gold bipyramid and silver nanorods purification and solvent exchange to CTAB***

To ensure proper coating of silver nanorods a CTAB induced depletion purification was necessary to separate the silver nanorods from other silver nanoparticle by-products.[15, 35] Moreover, a ligand exchange to CTAB was necessary for both gold bipyramids and silver nanorods to enable the further coating with silica. To separate AgNRs and AuBPs from nanoparticle by-products, a shape-selective purification procedure based on CTAB-induced depletion interactions was employed, adapted from previous reports. The as-synthesised nanoparticle dispersions were first washed three times with 0.5 mM CTAB solution to remove excess reactants and reduce the surfactant concentration below the critical micelle concentration. After each washing step, the nanoparticles were collected by centrifugation, the supernatant was discarded, and the pellet was redispersed in fresh 0.5 mM CTAB solution. Subsequently, the washed nanoparticles were redispersed in concentrated CTAB solutions to induce shape-selective sedimentation through depletion interactions. The CTAB concentration and incubation time were adjusted according to particle size: 0.09 M CTAB and approximately 12 h for AuBPs described in Figure 2 (L = 190.4 nm), and 0.10 M CTAB and approximately 20 h for smaller Ag NRs (L = 123.33 nm). The dispersions were maintained at 25–30 °C during incubation. At elevated CTAB concentrations, surfactant micelles induce depletion attractions between nanoparticles, preferentially promoting the aggregation and sedimentation of anisotropic particles due to their larger excluded-volume overlap compared with spherical by-products. Consequently, Au BPs and Ag NRs preferentially sedimented, while a substantial fraction of spherical and smaller nanoparticle impurities remained dispersed in the supernatant. Following incubation, the supernatant was carefully removed, and the sediment enriched in anisotropic nanoparticles was collected and washed three times with 0.5 mM CTAB solution to remove excess surfactant. Finally, the purified Au BPs and Ag NRs were redispersed in 0.5 mM CTAB solution for subsequent silica coating step.

***Synthesis of silica-coated gold-bipyramids and silver nanorods***

Following purification, gold bipyramids and silver nanorods were coated according to the same protocol as gold nanorods. Precursor amounts were adjusted according to nanoparticle geometry and concentration. For each coating reaction, a nanoparticle dispersion containing approximately $4.05 \times 10^{11}$ particles was prepared in 700 µL of 0.5 mM CTAB solution. For Au BPs (L=190.4 nm, W=63.5 nm, estimated surface area of 19,030 $nm^2$ per particle), 4 mL of the stock dispersion in 10 mM CTAB was used, whereas for AgNRs (L=123.33 nm, D=33.74 nm, estimated surface area of 16,652 $nm^2$ per particle), 1.536 mL of the stock dispersion in 10 mM CTAB was used. Both dispersions were washed three times with 0.5 mM CTAB by centrifugation at 4500 rpm (1811 g, Universal 320, Hettlich Zentrifugen) for 10 minutes and finally redispersed in 700 µL of 0.5 mM CTAB. An aliquot of a 100 mM ethanolic MPTMS solution was added, under stirring at 400 rpm. After stirring for 3 hours, an aliquot of a 20 vol% TEOS solution in methanol was added under stirring at 800 rpm and stirred for 1 hour. The precursor volumes were adjusted using particle-specific scaling factors. For AuBPs, 5.74 × 1.34 µL of 100 mM MPTMS and 17.36 × 1.34 µL of 20 vol% TEOS were added (scaling factor = 1.34), targeting a silica shell thickness of approximately 2 nm. Similarly, for Ag NRs, 5.74 × 2.885 µL of 100 mM MPTMS and 17.36 × 2.885 µL of 20 vol% TEOS were added (scaling factor = 2.885), targeting a shell thickness of approximately 6 nm. After TEOS addition, 8 µL of 0.01 M NaOH were added under stirring at 400 rpm. After 20 min, the reaction was left undisturbed for approximately 20 h at room temperature. The silica-coated nanoparticles were then purified by dilution with DMF and centrifugation at 9000 rpm (8960 rcf, Universal 320, Hettlich Zentrifugen ) for 15 min. The washing procedure was repeated three times, and the final pellet was redispersed in 1ml DMF.

***Scanning Trasmission Electron Microscopy (STEM) and Electron Energy-Loss Spectroscopy***

Transmission electron microscopy (TEM), scanning transmission electron microscopy (STEM) images were acquired using a FEI Talos F200X microscope operated at 200 kV, equipped with a Super X EDX spectroscopy detector (20kV range). High-angle annular dark-field (HAADF), bright-field (BF), and low angles dark-field (DF) detectors were employed for imaging. The sample was placed on a TFS low-background, double-tilt holder allowing for ±30° tilt. STEM-EELS measurements were performed in the spectral image (SI) mode in a JEM F200 operated in STEM-mode at 80kV (chosen to reduce sample damage and Cherenkov/retardation losses in $SiO_2$) equipped with a high-resolution EELS spectrometer Continuum S of GATAN (ZLP full width at half maximum 0.34 eV, with 2.5mm aperture, 15 meV/channel β 50.0 mrad, α 6.0 mrad) and acquired with the corresponding annular dark field (ADF) images. The sample was placed on a JEOL single tilt holder. The samples were prepared by depositing 10 µL of sample onto a copper Lacey grid (Ø 3,05 mm, Plano GmbH), and the grids were left to dry in air for at least one day. Prior to deposition the samples were purified by solvent exchange with MilliQ water to fully remove surfactants and DMF. Shortly, each sample was centrifuged twice at either 9000 rpm (8960 rcf) for AuNRs or 4500 rpm (1811 rcf) for 10 min and the surnatant was substituted by water each time. Analysis of the nanoparticle sizes was carried out using the program ImageJ, averaging on 100 particles.

***UV-vis-NIR spectroscopy***

UV-vis-NIR measurements were taken after every synthesis and purification. The spectra were recorded on a Jasco V-670 UV-vis/NIR spectrophotometer with following conditions: scan speed 400 nm/min, UV/Vis band width 0.5 nm, NIR band width 8.0 nm. The measurements were conducted with a 2 mm quartz cuvette from Hellma Analytics (Art. No. 100-5-40). Samples of uncoated nanoparticles had concentrations in the range 80-120 mM and were diluted 1:80-1:120 with ultrapure water before each measurement. Coated nanoparticles solutions before and after purification were analyzed without further dilution, using the same parameters and cuvette.

***Surface integral equation (SIE) simulations***

In the nonrecoil approximation, the small fractional change in the electron energy and momentum is neglected, and the electron follows the prescribed constant-velocity, straight-line trajectory

$$\mathbf{r}_e(t) = \left(b_x, b_y, vt\right) = (\mathbf{b}, vt),$$

where $\mathbf{b}$ is the impact parameter in the plane perpendicular to the electron trajectory. The interaction with the sample produces an induced electric field $\mathbf{E}_{\text{ind}}$, which performs work on the electron. The spectral loss probability is given by

$$\Gamma(\mathbf{b},\omega) = \frac{e}{\pi\hbar\omega}\int_{-\infty}^{\infty} d\,t\,\mathrm{Re}\big[e^{-i\omega t}\mathbf{v}\cdot\mathbf{E}_{\mathrm{ind}}(\mathbf{r}_e(t),\omega)\big],$$

where $e > 0$ is the magnitude of the electron charge.[25, 36] The corresponding mean energy loss is

$$\Delta E(\mathbf{b}) = \int_{0}^{\infty} d\,\omega\,\hbar\omega\,\Gamma(\mathbf{b},\omega).$$

Using the induced electromagnetic Green tensor and the convention of Garcia de Abajo and Kociak,[25] the spectral loss probability can equivalently be written as

$$\Gamma(\mathbf{b},\omega) = -\frac{4e^2}{\hbar}\int dz\,dz'\,\mathrm{Im}\big[e^{i\omega(z\prime-z)/v}G_{zz}^{\mathrm{ind}}\big((\mathbf{b},z),(\mathbf{b},z'),\omega\big)\big].$$

Here, $G^{\mathrm{ind}} = G - G^{(0)}$ is the induced (scattered) Green tensor.

We evaluate this expression using a discrete set of $z$-oriented electric-dipole sources placed along the electron trajectory.[32] For each source position $z'$, we evaluate the induced longitudinal electric field at observation positions $z$ along the same trajectory. The contributions from all source and observation positions are combined coherently with the phase factor $\exp[i\omega(z'-z)/v]$. The trajectory length and spatial sampling are increased until the calculated spectra are converged. For the BEM calculations, we use an in-house software package based on a surface-integral equation formulation of Maxwell's equations.[37] The particle surface is discretized into triangular elements, on which the surface-integral equations are solved for the induced surface currents. The dielectric functions of gold and silver are taken from Johnson and Christy.[38] The particle dimensions are obtained from the experimental measurements, and the mesh is refined until the calculated spectra are converged. For comparison with the experimental data, each calculated EELS map is normalized to its maximum intensity. The electron trajectory is discretized into 100 dipole sources after verifying convergence, and its total length is set to 2000 nm (from -1000 nm to 1000 nm) so that the induced field has decayed sufficiently at both ends of the integration interval.

Alternatively, the electron field can be introduced directly through the frequency-domain potentials of a uniformly moving point charge, as implemented in BEM formulations of EELS.[33, 39] This provides an alternative numerical evaluation of the same loss probability.

The kinetic energy $T$ of the electron is related to its relativistic speed $v$ by $T = (\gamma - 1)m_e c^2$, with $\gamma = 1/\sqrt{1 - v^2/c^2}$. Solving for the speed gives

$$v = c\sqrt{1 - \left(1 + \frac{T}{m_e c^2}\right)^{-2}},$$

where $m_e c^2 = 511$ keV is the electron rest energy. This gives $\frac{v}{c}$ = 0.446 − 0.777 for typical kinetic energies of 60 − 300 keV. For the simulations reported here, we use the experimental beam energy $T$ = 80 keV, corresponding to $\frac{v}{c}$ = 0.5.

***Fluorescence enhancement calculations***

To calculate the modification of spontaneous emission, we use dimethylformamide (DMF) as the surrounding solvent, in which the dyes and shelled plasmonic nanoparticles are dispersed. The total decay rate of an electric-dipole emitter is proportional to the dipole-projected local density of optical states at the emitter position and emission frequency. Because the total decay rate contains both radiative and nonradiative contributions, we evaluate these contributions separately. The radiated power is obtained by integrating the outward time-averaged Poynting flux over a closed surface enclosing the emitter and nanoparticle at a large distance. The power absorbed by the nanoparticle is obtained by integrating the Ohmic dissipation density over the particle volume.

The quantum yield of the emitter-nanoparticle system is:

$$Q = \frac{\mathit{\Gamma}_{rad}}{\mathit{\Gamma}_{rad} + \mathit{\Gamma}_{nrad} + \gamma_{nr}},$$

where $\mathit{\Gamma}_{rad}$ is the radiative decay rate in the presence of the nanoparticle, $\mathit{\Gamma}_{nrad}$ is the nanoparticle-induced nonradiative decay rate and $\gamma_{nr}$ is the intrinsic nonradiative decay rate of the dye.

Let $q = \gamma_{rad}/(\gamma_{rad} + \gamma_{nr})$ be the intrinsic quantum yield of the dye in homogeneous DMF. Defining $F_{rad} = P_{rad}/P_0 = \mathit{\Gamma}_{rad}/\gamma_{rad}$ and $F_{nrad} = P_{abs}/P_0 = \mathit{\Gamma}_{nrad}/\gamma_{rad}$, where $P_0$ is the power radiated by the same dipole in homogeneous DMF, the modified quantum yield becomes:

$$Q = \frac{F_{rad}}{F_{rad} + F_{nrad} + q^{-1} - 1}$$

The modified quantum yield depends on the intrinsic quantum yield of the dye and on the nanoparticle material and geometry, the emission wavelength, the emitter position and orientation and its separation from the particle. Emitters with moderate or low intrinsic quantum yield can exhibit a particularly large relative increase in quantum yield when radiative-rate enhancement dominates over nanoparticle-induced absorption. At short separations, however, nonradiative energy transfer to the metal can quench the emission.

Under weak excitation and assuming unchanged collection efficiency, the fluorescence-enhancement factor can be written as:

$$\eta_{fl} = \eta_{exc} \frac{Q}{q}$$

where $\eta_{exc}$ is the excitation-rate enhancement at the absorption wavelength. If the nanoparticle also changes the emission pattern, a collection-efficiency factor must [4]additionally be included.

**Aknowledgements**

The authors thank Guillermo González-Rubio and Ana Sánchez-Iglesias for their valuable feedback with the synthesis of AuNRs and AuBPs, Georg Haberfeldner for the discussions on STEM-EELS, and Fabian Gertsch and Junxi Wang for their help with the nanoparticle synthesis. The authors gratefully acknowledge the financial support by the Swiss National Science Foundation (Project No. CRSII5_216629) and by ETH Zürich. The authors also thank the Scientific Center for Optical and Electron Microscopy (ScopeM) of ETH Zürich for providing access to the electron microscopy facilities.

**Statement of contributions**

EGR: conceptualization (lead), formal analysis, investigation, methodology, supervision of students, writing – original draft (lead)

SA: formal analysis (lead), methodology, writing – original draft

RES: data acquisition, resources, writing – review & editing

OJFM: funding acquisition, project administration, resources, supervision, writing – review & editing

MN: funding acquisition, project administration, resources, supervision, writing – review & editing

FK: conceptualization, investigation, methodology, supervision of students, writing – review & editing